\documentclass[aps, twocolumn]{revtex4}
\usepackage{bm}
\usepackage{amsmath}
\usepackage{upgreek}
\usepackage{hyperref, cleveref}
\usepackage[english]{babel}
\usepackage[utf8]{inputenc}
\usepackage[colorinlistoftodos, color=green!40, prependcaption]{todonotes}
\begin{document}

\title{Study of Particle Production in Au+Au Collisions
at $\sqrt{s_{NN}} =$ 11.5-200 GeV Using HYDJET++ Framework}

\author{Gauri Devi$^{1}$\footnote{gauri.devi13@bhu.ac.in; gauri1995devi@gmail.com},\\ Satya Ranjan Nayak $^{1}$\footnote{satyanayak@bhu.ac.in}\\ and B. K. Singh$^1$\footnote{bksingh@bhu.ac.in; director@iiitdmj.ac.in(Corresponding author)}$^{,2}$}
 \affiliation{$^{1}$ Department of Physics, Institute of Science, Banaras Hindu University (BHU), Varanasi 221005, India  \\ $^{2}$ Discipline of Natural Sciences, PDPM Indian Institute of Information Technology Design \& Manufacturing, Jabalpur 482005, India }


\begin{abstract}
\noindent
 Using the HYDJET++ model, we study the production of multi-strange particles($\phi$ and $\Omega$) in Au+Au collisions at $\sqrt{s_{NN}} =$ 11.5, 19.6, 27, 39, and 200 GeV as functions of transverse momentum and centrality. The model calculations employ earlier freeze-out hypersurfaces for multi-strange particle production and are compared with STAR experimental data. Results indicate that the HYDJET++ model demonstrates a better agreement with experimental data for multi-strange particles at higher energies, particularly in relation to early thermal freeze-out. Meanwhile, the default version performs more accurately for lighter, non-strange particles at the same energy scales. We also analyze identified particle and mixed particle ratios, providing insights into strangeness enhancement. The results provide additional constraints on parton energy loss models, contributing to a more precise determination of the transport properties of hot and dense QCD matter.
\end{abstract}

\maketitle
\section{Introduction}
\label{Introduction}

In high-energy relativistic nuclear collisions, extreme temperature ($150\sim 170$ MeV) and energy densities ($ 1-5$ \text{GeV/$fm^{3}$}) are achieved~\cite{PHENIX:2004vcz}, which leads to the formation of a free state of quarks and gluons known as the Quark-Gluon Plasma (QGP). As the system expands and cools, it transitions back into hadronic matter, during which a large number of secondary particles are produced. By analyzing these particles, deeper insights into the thermodynamic properties of the system, like temperature, entropy density, energy density, and particle multiplicity can be obtained~\cite{PHENIX:2004vcz}. Particularly, at lower collision energies, these observables show distinctive changes that reflect the system's inability to sustain QGP formation and reveal important information about the Quantum Chromodynamics (QCD) phase diagram and the onset of deconfinement~\cite{STAR:2017sal}. The QCD phase diagram is typically represented as a temperature graph ($T$) against baryon chemical potential ($\mu_B$). $\mu_{B}$ plays an important role, reflecting the abundance of baryons over antibaryons and serving as a crucial statistical thermodynamic parameter controlling the net baryon density of the hadronic fireball system. A thermalized system of different $T$ and $\mu_B$ can be created in heavy-ion collisions by adjusting the collision energy ($\sqrt{s_{NN}}$) ~\cite{STAR:2008inc,STAR:2017sal,Ye:2017ooc, Shor:1984ui, Singh:1997ff, Singh:1986ia}.

The Relativistic Heavy Ion Collider (RHIC) took an experimental initiative called Beam Energy Scan (BES) to study the production and evolution of QGP at different beam energies and explore different phases of QGP~\cite{STAR:2008inc,STAR:2017sal,Ye:2017ooc}. In such an experiment, the energy dependence of $p_T$ spectra and ratios are crucial to understand the QGP formation~\cite{STAR:2017sal}. The particle spectra provide significant information about the system's temperature and collective flow, reflecting the dynamics of kinetic freeze-out.\\ The Solenoidal Tracker at RHIC (STAR)~\cite{STAR:2019bjj,STAR:2022tfp} and the Pioneering High Energy Nuclear Interaction eXperiment (PHENIX)~\cite{PHENIX:2013kod,PHENIX:2011rvu} at the RHIC experiments are designed to investigate relativistic nuclear collisions using a broad range of probes, with a particular emphasis on particles produced during the early stages of the collision, especially (multi-) strange hadrons~\cite{STAR:2019bjj,STAR:2022tfp}. 

In p+p collisions~\cite{PHENIX:2011rvu} hard-scattered partons fragment into jets of hadrons, these fragments are the primary source of hadrons, typically above $\geq$ 2 GeV/c. These p+p collisions serve as a crucial baseline for understanding nuclear effects and strangeness enhancement in heavy-ion (A+A) collisions~\cite{PHENIX:2011rvu}.

Among the various messengers, strange hadrons particularly the $\phi$ meson serve as an excellent messenger for exploring the QCD phase boundary and identifying the onset of deconfinement and critical phase conditions. It has two important features: first, these productions and decay resonances are affected by Okubo-Zweig-Iizuka (OZI) rule ~\cite{Shor:1984ui, Okubo:1963fa}. This rule strongly suppresses $\phi$-meson decay via three pions ($\phi\xrightarrow{} \pi^{+}+\pi^{-}+\pi^{0}$) while it does not suppress decay via two kaons ($\phi\xrightarrow{}K^{+}+K^{-}$ or $\phi\xrightarrow{}K_{L}^{0}+K_{S}^{0}$). Kaons decay gives a significant contribution (i.e., through first decay process, 49\% and second decay process, 34\%) in $\phi$-meson decay, and pions decay gives no significant contribution (i.e., 8\%) in $\phi$-meson decay and other's decay is not substantial. Second, their cross-section is small for scattering with non-strange hadrons during the hadronic phase~\cite{Shor:1984ui, Okubo:1963fa}.

In previous studies~\cite{Devi:2024ecm,Devi:2024cxy,Devi:2023wih,Singh:2023bzm}, the HYDJET++ model has been successfully utilized to describe strange particle production in the high-temperature and low–baryon chemical potential regime, corresponding to ultra-relativistic collision energies.
Now, we are exploring the low temperature and high baryon chemical potential regime, corresponding to lower collision energies by using the HYDJET++ model. This exploration provides new insights into energy dependence of particle production, strangeness enhancement, and QCD critical point through transverse momentum ($p_T$)-spectra, average transverse momentum ($\langle p_T \rangle$), $p_{T}$-spectra ratios with respect to the system's energies, baryon-to-meson, and multi-strange to single-strange hadron ratios.

The outline of this study is as follows: Section~\ref{model} provides a brief description of the HYDJET++model~\cite{Lokhtin:2012re,Lokhtin:2009hs,Lokhtin:2008xi,Lokhtin:2005px,Amelin:2006qe,Amelin:2007ic,Devi:2024cxy,Devi:2024uis,Singh:2023bzm,Bravina:2020sbz,Devi:2023wih}. Section~\ref{Results} A highlights the difference between the thermal freeze-out $T_{th}$ used in HYDJET++ and $T_{th}$s extracted using other methods like Tsallis, Blast-wave and exponential fits. Section~\ref{Results} B,  highlights the precise measurements of the $p_T$ spectra and average $\langle p_T \rangle$ for $\phi$ and $\Omega$ hadrons. These results are compared to the available STAR experiment data~\cite{STAR:2008bgi, STAR:2015vvs,STAR:2019bjj,STAR:2006egk}, focusing on collisions at beam scan RHIC energies~\cite{STAR:2008bgi}. Section~\ref{Results} C discusses the particle ratios in the RHIC Beam Energy Scan, while Section~\ref{conclusion} provides a summary of our main findings. 

\section{ Brief introduction to the HYDJET++ Model}
\label{model}
The HYDJET++ model combines the soft and hard components particle production simultaneously allowing the study of the dynamics of relativistic heavy-ion collisions\cite{Lokhtin:2005px, Lokhtin:2009hs}. 
The soft component in HYDJET++ denotes a thermalized hadronic state, simulated through the multiparticle hadron production using generator FAST MC~\cite{Amelin:2006qe, Amelin:2007ic}. This generator uses a parameterized relativistic hydrodynamics model that consider the chemical ($T_{\mathrm{ch}}$) and thermal ($T_{\mathrm{th}}$) freeze-out hypersurfaces to explain the system's evolution. At chemical freeze-out, hadrons are considered to be in chemical equilibrium, which determines the final particle composition. At thermal freeze-out, local thermal equilibrium is considered, which determines the momentum distributions of hadrons~\cite{Bravina:2017rkp}.

This stage involves statistical particle production from the grand canonical ensemble that incorporates thermodynamic parameters such as T, $\mu$, and system volume (V) and simultaneously includes treatment of collective flow effects to account for the medium's expansion. Here, the model assumes longitudinal boost invariance and incorporates transverse radial flow, quantified by the maximum transverse flow rapidity ($\rho_u^{\mathrm{max}}$)~\cite{Lokhtin:2008xi}. These flow effects significantly influence the shape of transverse momentum ($p_T$) spectra, especially for heavier particles, which receive a stronger boost during the medium's expansion.

On the other hand, the hard component in HYDJET++ involves to generate multi-jets using a binomial distribution inside the PYQUEN energy loss model~\cite{Lokhtin:2005px}. The PYQUEN model for single hard nucleon-nucleon subcollisions was developed as a modification of jet events produced by the PYTHIA\_6.4~\cite{Sjostrand:2006za} generator for hadron-hadron interactions~\cite{Lokhtin:2009hs}. Following this, the final hadronization of hard partons takes place according to the Lund string model~\cite{Lokhtin:2009hs}. The average number of jets produced in an AA collisions is determined by multiplying the binary NN collisions at a given impact parameter with the integral cross-section of the hard process in NN collisions with a minimum transverse momentum transfer ($p_{T}^{min}$). The $p_{T}^{min}$ serves as the threshold separating soft particle emission from hard scatterings associated with jet formation effects~\cite{Lokhtin:2005px,STAR:2017sal}. 

The production of heavier hadrons (such as strange and multi-strange baryons) is more sensitive to the freeze-out temperature due to their larger mass, requiring higher thermal energy for their creation. In contrast, lighter hadrons (such as pions) are more abundantly produced even at lower temperatures. Thus, the relative abundances and spectra of different particle species carry imprints of the system’s temperature and evolution. 

This version of HYDJET++ is an extended adaptation of the previous HYDJET++ implementation at RHIC and LHC energies~\cite{Devi:2024uis,Devi:2024cxy}, now tailored for Beam Energy Scan at RHIC energies. We have modified hadron-species dependent freeze-out parameters like $T_{th}$, $T_{ch}$, and a phase-space occupancy factor, $\gamma_s$~\cite{Castorina:2016eyx} in the default HYDJET++ framework. These parameters were fitted to experimental data~\cite{STAR:2017sal} based on the comparison of $p_{T}-$ spectra of charged yields between the modified and default HYDJET++ models~\cite{Devi:2024uis}. The last parameter $\gamma_s$ has been adjusted as a centrality-dependent variable~\cite{STAR:2008bgi}, as detailed in section~\ref{Results}. $\gamma_s$, is introduced to account for the potential nonequilibrium of strange quarks. \\
\begin{figure*}
    \includegraphics[width=0.6\linewidth]{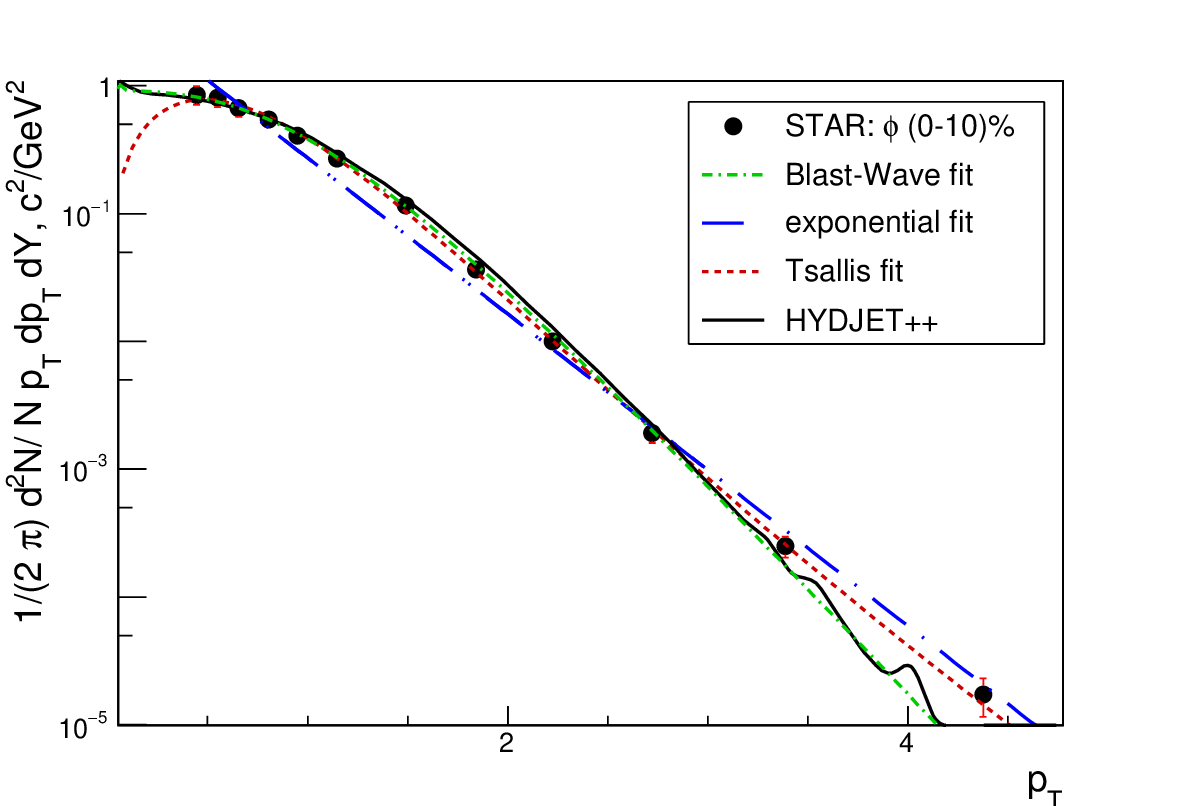}
    \caption{The $p_T$ spectra of $\phi$ meson using Tsallis distribution, Blast-wave, exponential fits and HYDJET++ for central collisions of Au+Au collisions at $\sqrt{s_{NN}}$=39 GeV. The markers denote the STAR experimental data~\cite{STAR:2019bjj}, while the lines correspond to the results obtained from different methods.}\label{fig1*}
   \end{figure*}
  \begin{table*}
     \centering
     
     \begin{tabular}{|c|c|c|c|c|c|c|c|c|c|c|}
     \hline
      \textbf{Energies (GeV)} & \multicolumn{2}{c|}{\textbf{11.5}} & \multicolumn{2}{c|}{\textbf{19.6}} & \multicolumn{2}{c|}{\textbf{27}} & \multicolumn{2}{c|}{\textbf{39}} & \multicolumn{2}{c|}{\textbf{200}} \\
    \hline
    \textbf{Centrality (\%)} & $\langle N_{part} \rangle$ & $\langle N_{coll} \rangle$ & $\langle N_{part} \rangle$ & $\langle N_{coll} \rangle$ & $\langle N_{part} \rangle$ & $\langle N_{coll} \rangle$ & $\langle N_{part} \rangle$ & $\langle N_{coll} \rangle$ & $\langle N_{part} \rangle$ & $\langle N_{coll} \rangle$ \\
   
         \hline
         0-10 &306.7&669.3 &307.4&683.4 &308.5&700.6  & 310&726.3 & 346.3&1006 \\
         \hline
        10-20  &212.8 & 420.5 & 213.6 & 429.3 & 214.7 & 440.2 & 216.1 & 456.1 &223.9&562.9\\
        \hline
        20-30  &145.4&257.8  &146.2&263.4  & 147.0 & 269.8 &148.2&279.6& 155.1&345.1 \\
        \hline
        30-40  & 95.20&148.6 &95.80&151.8 &96.51&155.6 &97.50&161.3 & 103.3&199.1 \\
      \hline
        40-60  & 46.35&59.27 &46.84&60.70 &47.28&62.18 &47.96&64.49 &--&-- \\
      \hline
      60-80  & 12.05&11.08 &12.20&11.32 &12.37&11.60 &12.63&12.03 & 14.25&14.97 \\
      \hline
      \end{tabular}
     \caption{Centrality-dependent values of the number of participants ($\langle N_{part} \rangle$), number of binary collisions ($\langle N_{coll} \rangle$), using HYDJET++ simulations for Au+Au collisions at $\sqrt{s_{NN}}$= 11.5-200 GeV.}\label{table1}
 \end{table*}
\begin{figure*}
    \centering
   \includegraphics[width=1.0\linewidth]{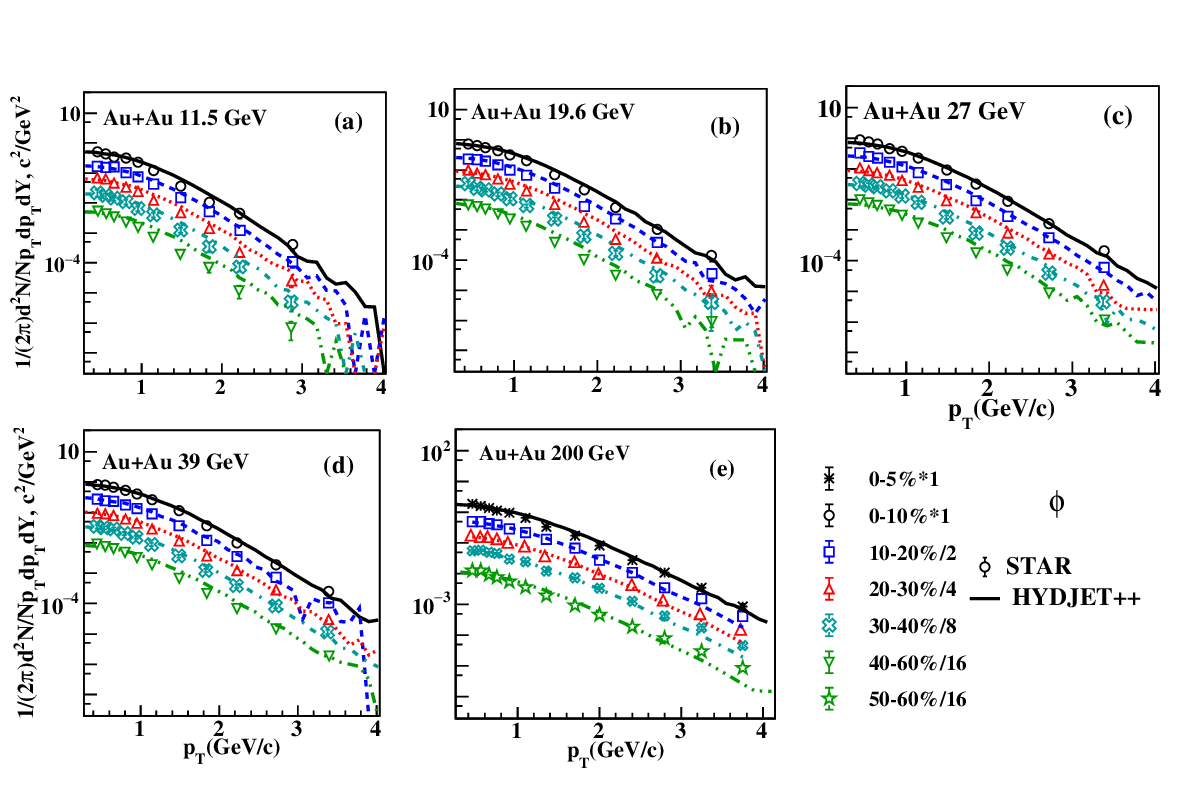}
    \includegraphics[width=1.0\linewidth]{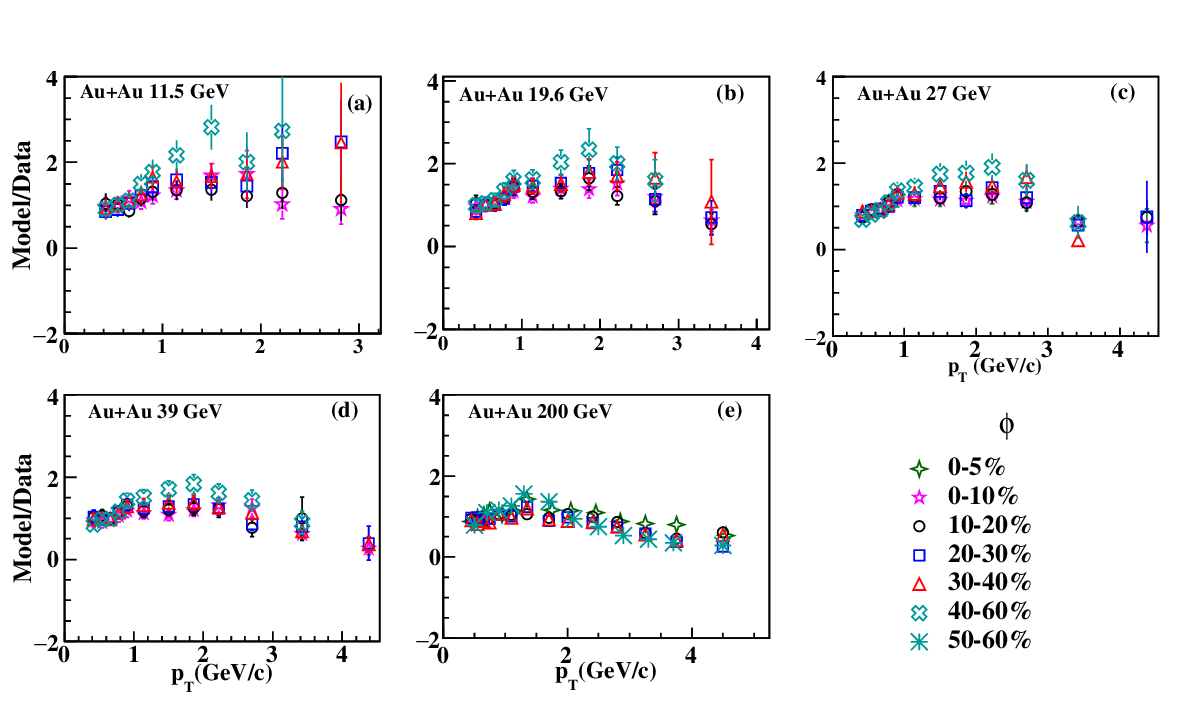}
    \caption{\textbf{Upper panel-} The $p_T$ spectra of $\phi$-meson as a function of $p_{T}$ in Au+Au collisions at $\sqrt{s_{NN}}= 11.5-200$ GeV. The lines represent HYDJET++ results while markers represent STAR experimental data~\cite{STAR:2008bgi, STAR:2015vvs,STAR:2019bjj}. \textbf{Lower panel-} Ratios of model predictions-to-experimental data for $\phi$ meson.}\label{fig2.1}
\end{figure*}
\begin{figure*}
    \centering
    \includegraphics[width=1.0\linewidth]{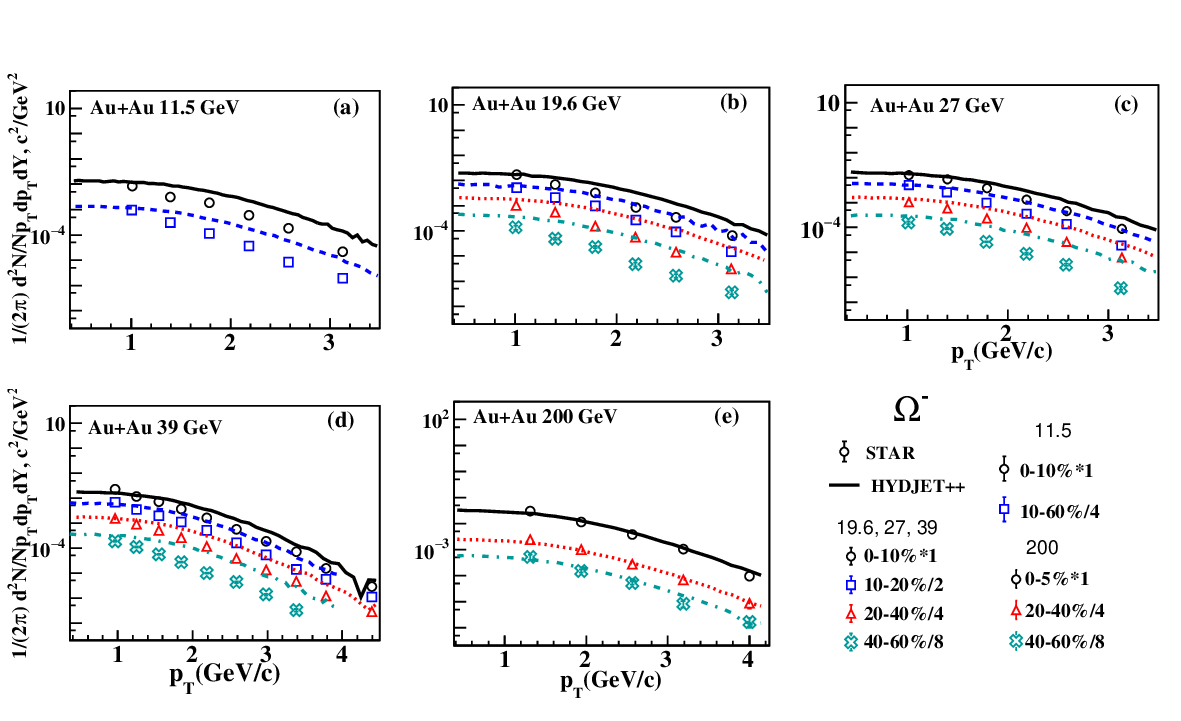}
     \includegraphics[width=1.0\linewidth]{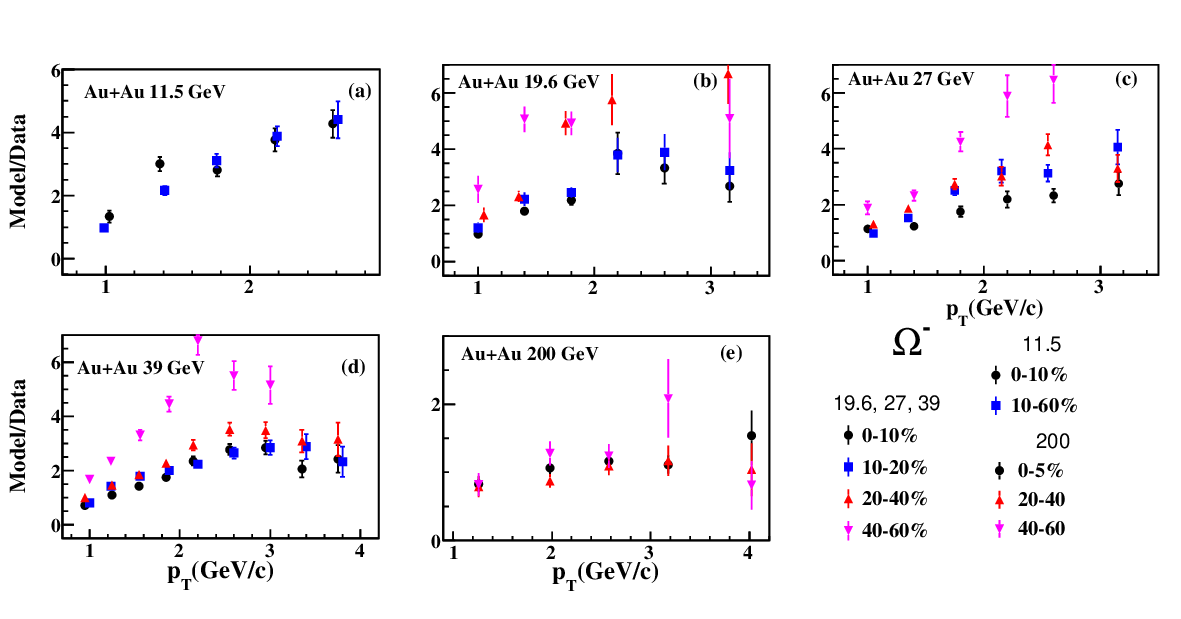}
    \caption{\textbf{Upper panel-} The $p_T$ spectra of $\Omega^{-}$-baryon  as a function of $p_{T}$ in Au+Au collisions at $\sqrt{s_{NN}}= 11.5-200$ GeV. The lines represent HYDJET++ results while markers represent STAR experimental data~\cite{STAR:2015vvs,STAR:2006egk,STAR:2019bjj}. \textbf{Lower panel-} Ratios of model predictions-to-experimental data for $\Omega^{-}$-baryon.}\label{fig2.2}
    \end{figure*}  
\begin{figure*}
    \centering
    \includegraphics[width=1.0\linewidth]{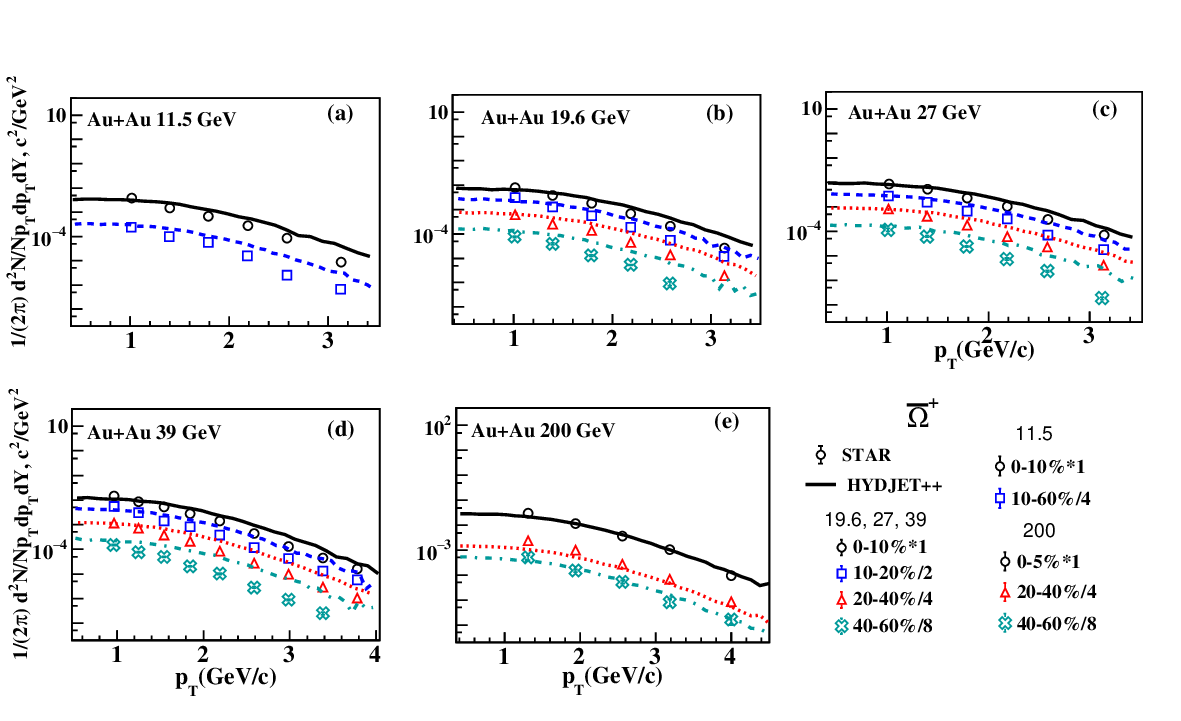}
       \includegraphics[width=1.0\linewidth]{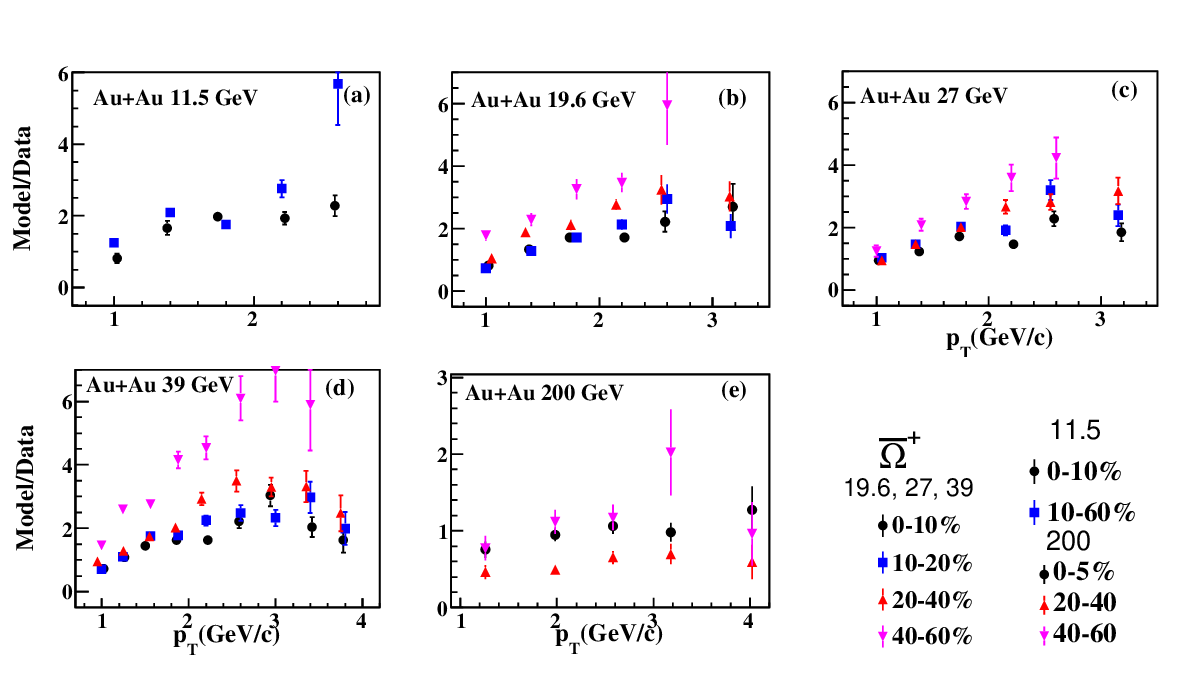}
    \caption{\textbf{Upper panel-} $p_T$ spectra of $\overline{\Omega}^{+}$-baryon from Au+Au collisions at $\sqrt{s_{NN}}= 11.5-200$ GeV. The lines represent HYDJET++ results while markers represent STAR experimental data~\cite{STAR:2015vvs,STAR:2006egk,STAR:2019bjj}. \textbf{Lower panel-} Ratios of model predictions-to-experimental data for $\overline{\Omega}^{+}$-baryon.}\label{fig2.3}
    \end{figure*}
 \begin{figure*}
    \centering
    \includegraphics[width=0.8\linewidth]{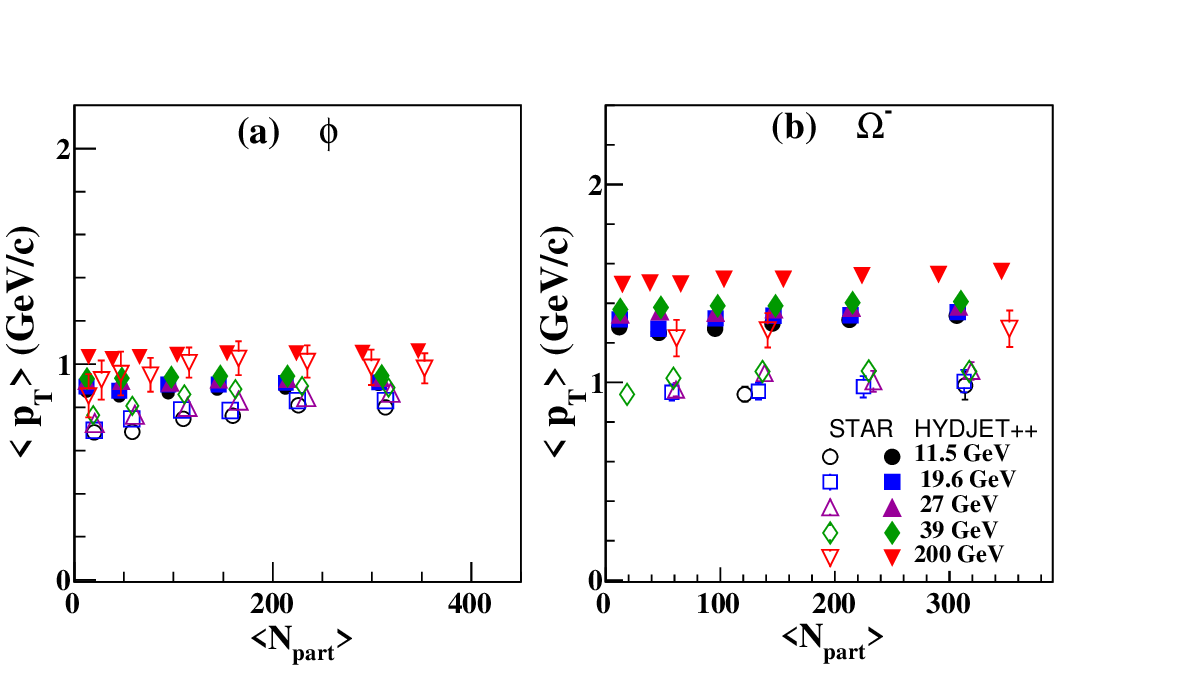}
    \caption{$\langle p_T \rangle$ as a function of $\langle N_{part} \rangle$ in Au+Au collisions at $\sqrt{s_{NN}}= 11.5-200$ GeV. The Solid markers represent HYDJET++ results while open markers represent STAR experimental data~\cite{STAR:2008med,STAR:2019bjj}}\label{fig2.4}
\end{figure*}
 \begin{figure*}
    \centering
    \includegraphics[width=1.0\linewidth]{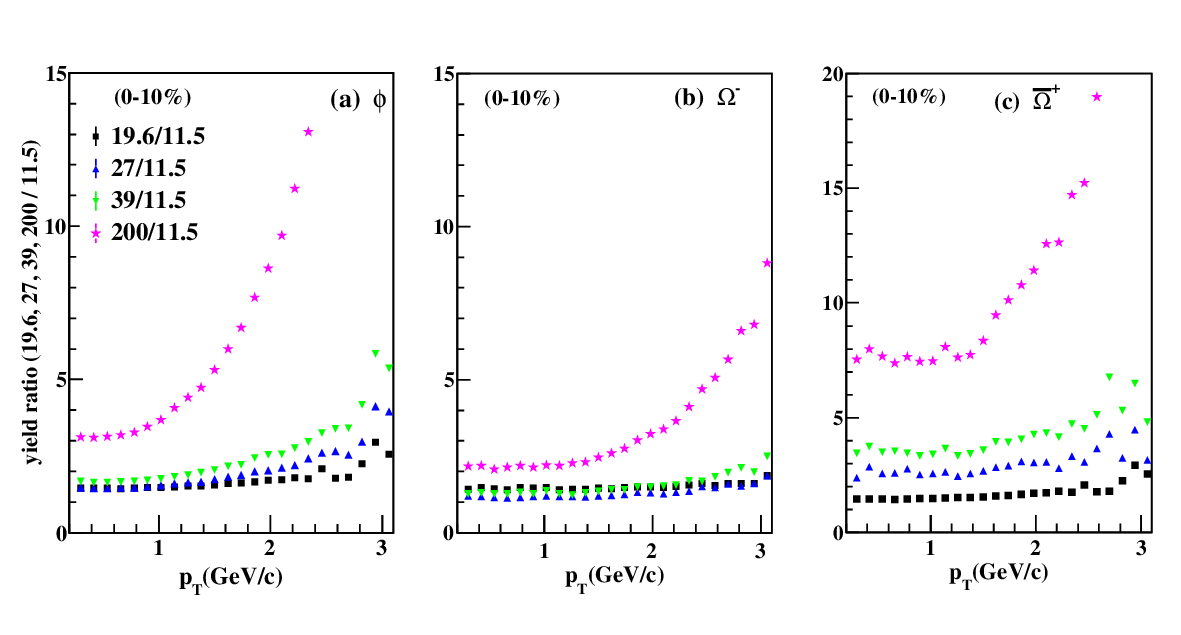}
    \caption{Particle yield ratios (in term of energy ratios) as a function of $p_T$ in Au+Au collisions at $\sqrt{s_{NN}}= 11.5-200$ GeV. The Solid markers represent HYDJET++ results.}\label{fig3}
   \end{figure*}
\begin{figure*}
    \centering
    \includegraphics[width=1.0\linewidth]{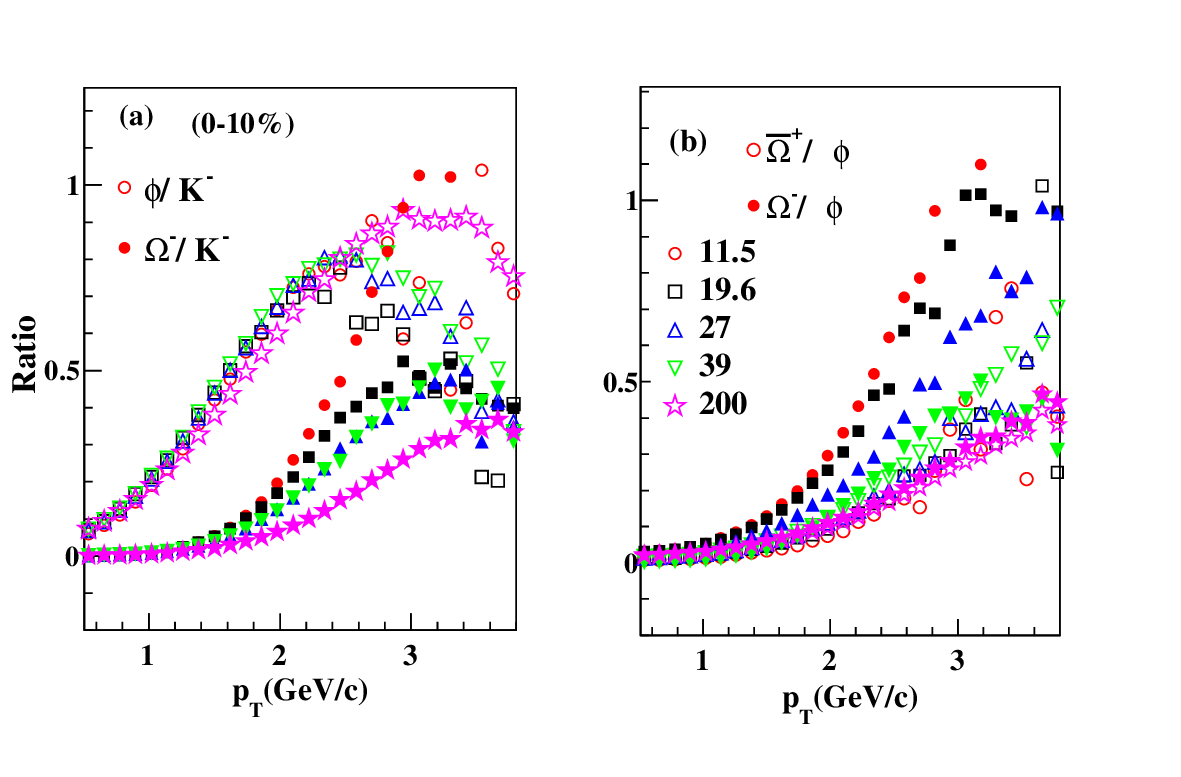}
    \caption{\textbf{(a)} Particle ratios, N($\phi)$/N($K^{-}$) (represents by Open markers), and N($\Omega^{-}$)/N($K^{-}$) (represents by Solid markers),  \textbf{(b)} N($\Omega^{-}(\overline{\Omega}^{+}))$/N($\phi)$) (represents by Solid (Open) markers), as a function of $p_{T}$ in Au+Au collisions at $\sqrt{s_{NN}}= 11.5-200$ GeV.}\label{fig3.1}
\end{figure*}
The geometrical parameters in HYDJET++, calculated using the 2 pF (two-parameter Fermi) form MC Glauber model, show consistency with the experimental data reported by the STAR collaborations~\cite{STAR:2017sal}, where the number of participating collisions $(\langle N_{part}\rangle)$ and the number of binary collisions $(\langle N_{coll}\rangle)$ are also obtained from Glauber calculations. In Table ~\ref{table1}, we present the correlation between the $\langle N_{coll}\rangle$ and $\langle N_{part} \rangle$ across various centrality classes ($\%$), as obtained from the geometrical output of the HYDJET++ model. Furthermore, by comparing the $p_T$ spectra to experiment data~\cite{STAR:2017sal}, we have employed the optimized chemical parameters ( $\gamma_s$, $\mu_{B}$, $T_{ch}$) and thermal parameter($T_{th}$, fitted using the blast-wave model~\cite{STAR:2017sal}) for RHIC beam scan energies , adopting the same values reported by the STAR experiment in
Reference~\cite{STAR:2017sal}. In STAR, these parameters were typically extracted through statistical thermal model analyses with the THERMUS package~\cite{Wheaton:2004qb},using either the grand-canonical ensemble (GCE) or the strangeness canonical ensemble (SCE). While GCE is suitable at higher energies with abundant strangeness and SCE is required at lower energies with limited strangeness production. In this work we adopt the GCE approach parameters, as the RHIC BES program spans a broad energy range.
\section{Results and Discussions}
\label{Results}
\subsection{Freeze-out temperatures}
The thermal freeze-out hypersurface determines the soft hadroproduction in HYDJET++, it is necessary to understand the difference between $T_{th}$ used in HYDJET++ and $T_{th}$s extracted using other methods. One of the popular ways to extract $T_{th}$ is to fit the data with a distribution like Tsallis or by the Blast-wave fit. A typical Tsallis distribution is given by~\cite{Tao:2022tcw}:

 \begin{eqnarray*}
\frac{d^2 N}{2\pi p_Tdp_Tdy}=gvm_T\frac{\cosh y}{(2\pi)^3}[1-(q-1)(m_T\cosh y-\mu)/T]
\label{eq:3}
\end{eqnarray*}
 where, $g$ is the degeneracy, $v$ is the volume, $T$ is same as $T_{th}$, and the non-extensive parameter $q$ refers to the system's deviation from thermalization, at $q=1$ it becomes Boltzmann's distribution. It was reported in a previous study that the value of $q$ increases with increasing beam energy, which is concerning as the systems formed at high $\sqrt{s_{NN}}$ are known to thermalize due to high temperature and longer lifetime of the QGP~\cite{Tao:2022tcw}. We have fitted the transverse momentum distribution of $\phi$ meson using the Tsallis distribution in figure~\ref{fig1*}. The resulting value of $T$ was 184 MeV, much higher than the $T_{th}$ used in this work. This difference arises as HYDJET++ assumes a fully thermal system while the Tsallis distribution assumes a near-equilibrated system.\\ 
In this work, we have matched the $p_T$ data by simultaneously tuning thermal freeze-out temperature $T_{th}$ and maximum transverse flow rapidity $\rho_{max}$. These parameters are similar to the ones used in Blast-wave fit. However, a key difference between these two is that the HYDJET++ is a Monte Carlo event generator rather than a fitting tool. Unlike the Blast wave model \cite{Schnedermann:1993ws}, the HYDJET++ simulations include contributions from jets and resonance decays, which affect the $p_T$ spectra and the choice of the $T_{th}$ parameter. Hence, the $T_{th}$ values used in this work are different than the ones extracted from the Blast wave fit from the corresponding paper \cite{STAR:2017sal}.

\subsection{Transverse momentum spectra ($p_T$) and mean $p_T$ ($\langle  p_{T} \rangle$) }
The transverse momentum spectra of the multi-strange particles are significant observables, since they provide essential information about the equilibrium dynamics and isotropy of the system in high-energy collisions. These calculations give the information that the production of $\phi$ and $\Omega$ supports the fact that jet production is suppressed in events where one or more strange particles are produced~\cite{Singh:2023bzm,Devi:2024cxy}.

\Cref{fig2.1,fig2.2,fig2.3} present the transverse momentum spectra of the $\phi$ meson and $\Omega$ baryons obtained from the HYDJET++ model in comparison with the STAR experimental data~\cite{STAR:2008bgi, STAR:2015vvs,STAR:2019bjj,STAR:2006egk}. The figures also present the model-to-data ratios, which indicate the level of agreement between the HYDJET++ predictions and the experimental results. The HYDJET++ calculations describe the data well in the central collision at higher energies than the peripheral collisions and lower energies. \Cref{fig2.2,fig2.3} show more uncertainty (approx 1.5 times in central and 4-6 times in peripheral collisions) with experimental data  across the entire $ p_T$ range as the collision energy decreases towards more central events. The uncertainty is higher for $\Omega$ baryon than the $\phi$ meson which strongly supports the flavor-dependence particle production at lower energies. These spectra also reveal a significant difference in the invariant yield values for particles and the corresponding antiparticles~\cite{STAR:2013ayu}. This difference increases with decreasing energy ($\sqrt{s_{NN}}$) (or increasing $\mu_B$ ) and is larger for the baryons. This also implies that particles and antiparticles are no longer consistent with the matter-antimatter symmetry towards lower energy that was observed at $\sqrt{s_{NN}}$ = 200 GeV~\cite{STAR:2013ayu}. 

\Cref{fig2.4} presents the mean $\langle p_{T}\rangle$ of both particles at RHIC BES energies and the available experimental results ~\cite{STAR:2008med,STAR:2019bjj}. \Cref{fig2.4}(a) presents that the results of the particles $\phi$ at RHIC BES energies match well within the uncertainty range for central collisions and higher energies. In contrast, \Cref{fig2.4}(b) shows that the HYDJET++ predictions for $\Omega$ baryons at RHIC BES energies overpredict the experimental data across all collision centralities and energies. This behavior suggests that the system may not reach full thermal equilibrium for $\Omega$ baryons at lower collision energies. The freeze-out conditions appear to be more accurately tuned for mesons like the $\phi$ than for baryons like the $\Omega$, leading to the observed discrepancies. Additionally, the model is also less sensitive to certain mechanisms, such as baryon coalescence which carries more weight for multi-strange baryons like $\Omega$~\cite{Waqas:2021bsy,Devi:2024uis}.
\begin{figure}
     \includegraphics[width=1.0\linewidth]{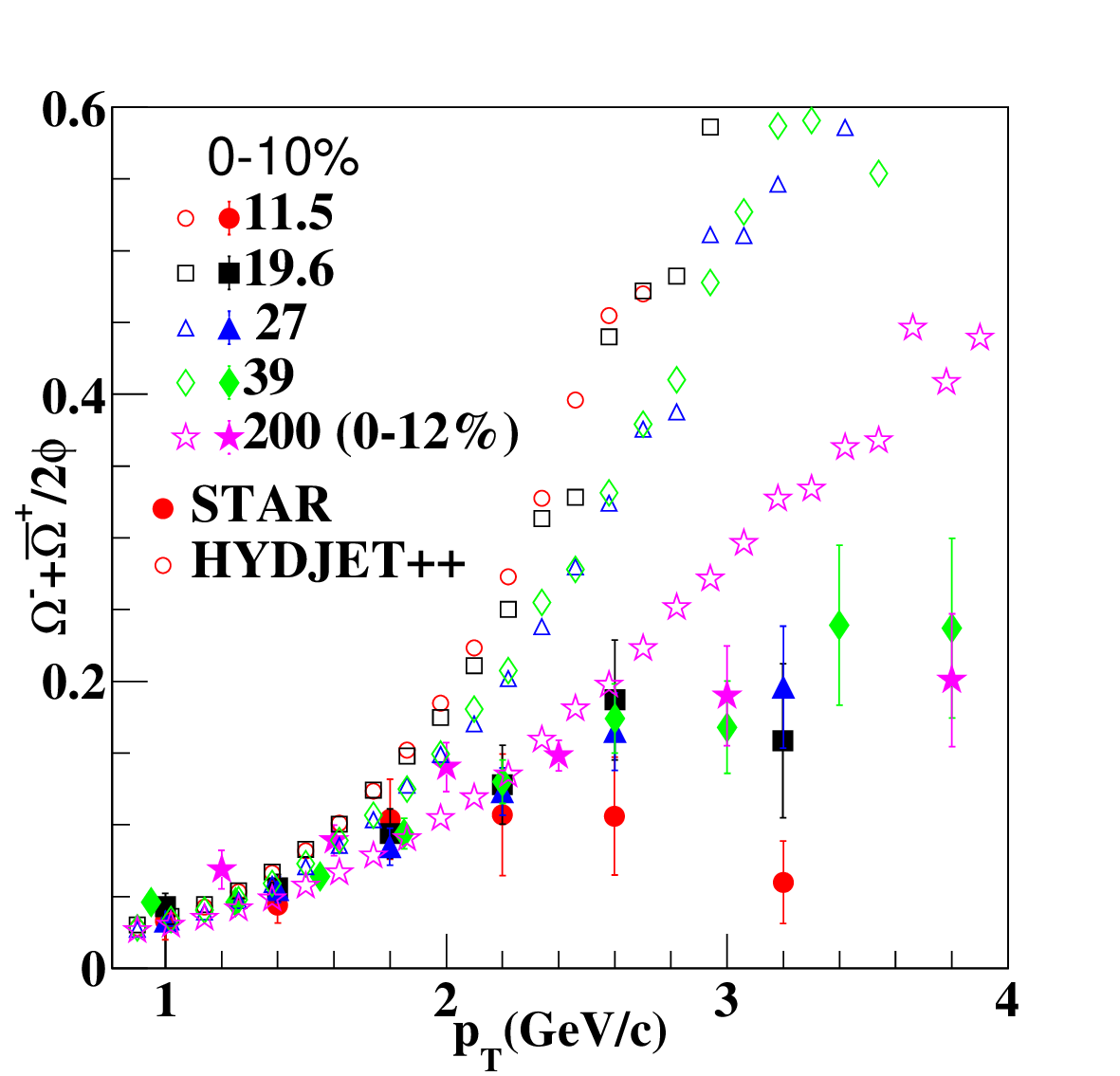}
    \caption{Baryon-to-meson ratio, N($\Omega^{-}+\overline{\Omega}^{+})$/[2N($\phi$)] as a function of $p_{T}$ in Au+Au collisions at $\sqrt{s_{NN}}= 11.5-200$ GeV. The Open markers represent HYDJET++ results and solid markers represent STAR experimental data~\cite{STAR:2015vvs}.}\label{fig4.2}
   \end{figure}

   \begin{figure}
    \includegraphics[width=1.0\linewidth]{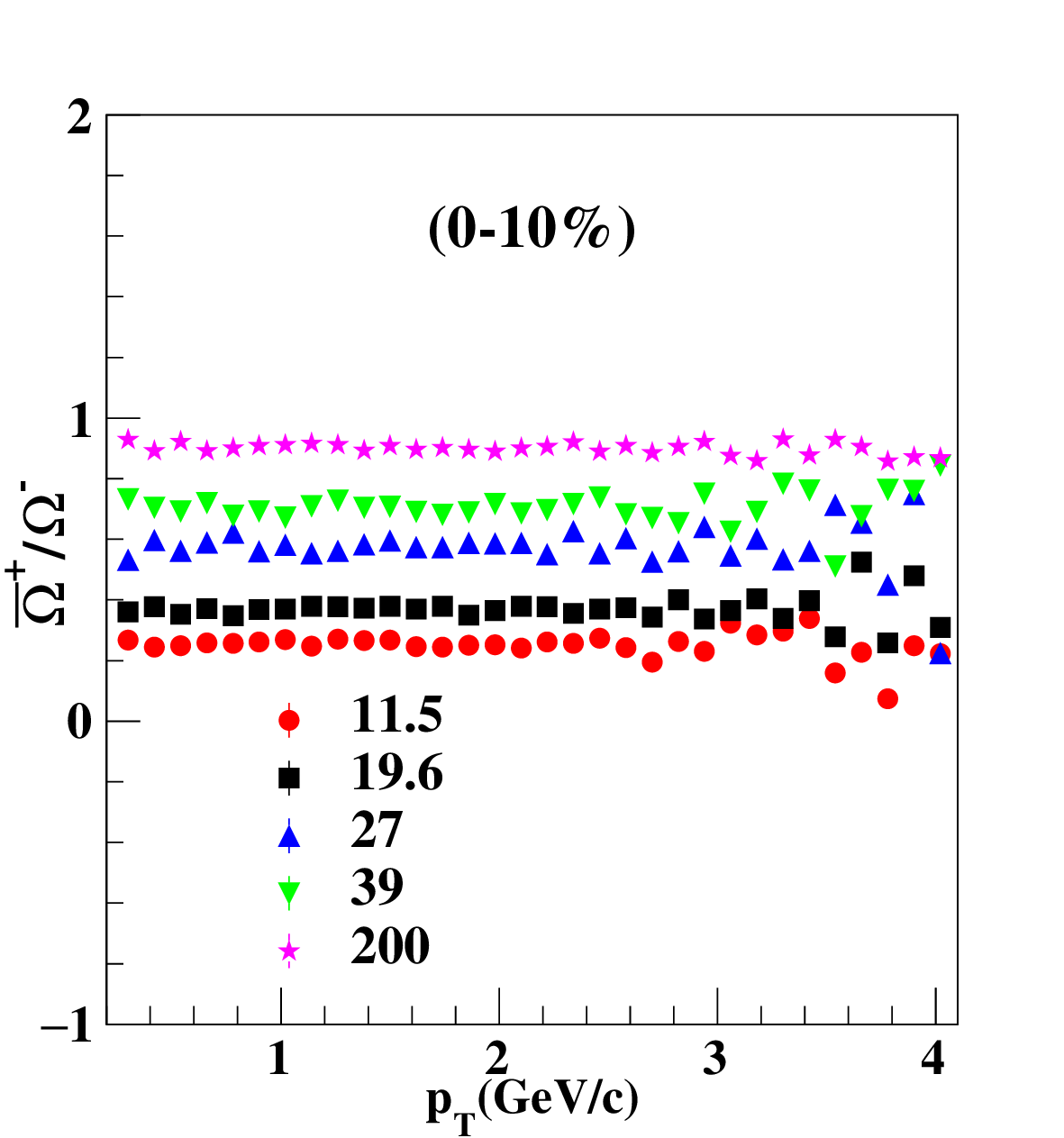}
    \caption{Antiparticle-to-particle ratio, $N(\overline{\Omega}^{+})/ N(\Omega^{-})$ as a function of $p_{T}$ in Au+Au collisions at $\sqrt{s_{NN}}= 11.5-200$ GeV. The Solid markers represent HYDJET++ results. }\label{fig4}
   \end{figure}
   
\subsection{Particle Ratio}

\Cref{fig3,fig3.1,fig4,fig4.2} present the particle yield ratios (in terms of energy ratios)  and different particle ratios, N($\Omega^{-}(\overline{\Omega}^{+}))$/[N($\phi$)], $N(\overline{\Omega}^{+})/ N(\Omega^{-})$, N($\phi(\Omega^{-}))$/N($K^{-}$), and N($\Omega^{-}+\overline{\Omega}^{+})$/[2N($\phi$)] as a function of $p_T$, and compare with the available experimental data~\cite{STAR:2015vvs,Aichelin:2010ed} for most central collisions from Au+Au collisions at $\sqrt{s_{NN}}= 11.5-200$ GeV. \\
\Cref{fig3} presents the particle yield ratios at 19.6, 27, 39, 200 GeV to 11.5 GeV, show a suppression of particle production with flavor towards high to low-energy systems while show an enhancement of particle production as a function of $p_T$. \\
\Cref{fig3.1}(a) presents the baryon-to-meson (N($\Omega^{-})$/N($K^{-}$)) and meson-to-meson (N($\phi)$/N($K^{-}$)) ratios. These results show different shapes of particle $p_T$ distributions at low and intermediate $p_T$ in central collisions, which indicates that the number of quarks is the main factor and baryons are suppressed more than mesons. \Cref{fig3.1} (b) presents the ratios $\Omega^{-}/\phi$ at various energies. At intermediate $p_{T}$ ($\sim 2-4 $ GeV/c), ratio of $\overline{\Omega}^{-}/\phi$ observed to be approximately 2–3 times higher than the ratio of $\overline{\Omega}^{+}/\phi$ with decreasing energy. This increase can be attributed to the reduction in $\overline{\Omega}^{+}$ baryon production compared to $\Omega^{-}$ baryons, caused by annihilation processes affecting antibaryon production in the baryon-rich QCD matter at lower energies~\cite{STAR:2019bjj}.\\
Expanding on this,  \Cref{fig4.2} presents the  N($\Omega^{-}+\overline{\Omega}^{+})$/[N($\phi$)] ratios in
central ($0-10\%$) Au+Au collisions at $\sqrt{s_{NN}}= 11.5-200$ GeV exhibit a consistent qualitative increase with $p_T$ displaying an approximately linear trend similar to the STAR experiment results~\cite{STAR:2015vvs}. However, the model systematically overpredicts the experimental values, particularly at lower collision energies. This overprediction may arise from the model’s treatment of strangeness enhancement, where the parametrization of strange quark production in the soft component may not adequately reflect the reduced partonic collectivity and incomplete thermalization at lower energy. \\
Furthermore, \Cref{fig4} presents the ratio of antiparticle-to-particle ($\overline{B}/B$) at $\sqrt{s_{NN}}= 11.5-200$ GeV. A variation with energy supports that the matter-antimatter symmetry does not hold at lower energies.  At lower energies, the ratio tends to be less than 1 which supports the notion that antiparticle production is less than particle production. This trend is consistent with experimental results for the RHIC beam scan energy~\cite{STAR:2019bjj}. 
This property of antiparticle/particle ratios is qualitatively well described in the statistical model~\cite{Cleymans:2006xj,Nayak:2024jbt}. 
 
\section{Conclusion}
\label{conclusion}
In conclusion, our study demonstrates that the simulated bulk properties using the HYDJET++ model are sensitive to the choice of input parameters. Details of the analysis method for $\phi$ meson and $\Omega$ baryons are presented. This study represents the dependence of multi-strange particle production on energy, $\langle N_{part}\rangle$, and $\langle N_{coll}\rangle$, along with the differences in freeze-out temperatures between HYDJET++ and other thermal extraction methods, is reported. Additionally, results on $p_T$ spectra, $\langle p_{T} \rangle$, particle yield ratios (relative to 11.5 GeV energies) as a function of $ p_{T}$, particle ratios for five collision systems are presented.\\ By comparing the model predictions with the experimental data, we understand the multi-strange particle production mechanism at RHIC beam energy scan. The $p_T$ spectra which illustrate that the hardening of spectra with increasing collision energy and a clear dependence on particle flavour. This is also suggest a growing collective radial flow and reflect the changing freeze-out conditions of the system. It also suggests the $\phi$ achieves the thermal resonance equilibrium while $\Omega$ does not achieve equilibrium due to heavier mass with less hadronic interaction.\\
Additionally, in calculations of particle yield ratios (relative to 11.5 GeV energies) of multi-strange hadrons at low to intermediate $p_T$ in central collisions, we have found a strong dependence of the $p_T$-shape in $\phi$ and $\overline{\Omega}^{+}$ than $\Omega^{-}$ towards $19.6/11.5$ to $200/11.5$ energy ratios.\\ Following this, the baryon-to-meson and meson-to-meson ratios show the suppression of strange-quark production in
11.5 GeV compared to $\sqrt{s_{NN}}=$ 19.6 GeV. These features suggest that there is likely a change in the underlying strange-quark dynamics in the bulk QCD matter responsible for $\Omega$ and $\phi$ production. Our measurements point to collision energies below 19.6 GeV for further investigation of a possible transition from partonic dominant
matter ($\sqrt{s_{NN}}>$ 19.6 GeV) to hadronic dominant matter ($\sqrt{s_{NN}}< $11.5 GeV).\\We also see that the ratios are closer to the value of one towards 200 GeV, reflecting the approach to baryon-antibaryon symmetry at midrapidity as the collision
energy increases.\\
Furthermore, by analyzing the $p_T$-spectra, and particle ratio patterns across various beam energies, we can provide more comprehensive insights into the evolution of the QGP medium and the mechanisms governing multi-strange particle production in heavy-ion collisions. These findings not only enhance our understanding of strangeness dynamics in a baryon-rich QCD region but also serve as a valuable reference for selecting optimal model parameters in future beam energy dependent studies.

\section{Acknowledgments}
BKS gratefully acknowledges the financial support provided by the BHU Institutions of Eminence (IoE) Grant No. 6031, Govt. of India. GD and SRN thank non-net fellowships under the central university scheme.

\end{document}